# Mechanisms of shock-induced initiation at micro-scale defects in energetic crystal-binder systems

*Pratik Das and H. S. Udaykumar\**

*Department of Mechanical and Industrial Engineering*

*The University of Iowa, Iowa City, IA-52242*

## Abstract

Crystals of energetic materials, such as 1,3,5,7-Tetranitro-1,3,5,7-tetrazocane (HMX), embedded in plastic binders are the building blocks of plastic-bonded explosives (PBX). Such heterogeneous energetic materials contain microstructural features such as sharp corners, interfaces between crystal and binder, intra- and extra-granular voids and other defects. Energy localization or "hotspots" arise during shock interaction with the microstructural heterogeneities, leading to initiation of PBXs. In this paper, high-resolution numerical simulations are performed to elucidate the mechanistic details of shock-induced initiation in a PBX; we examine four different mechanisms: 1) Shock-focusing at sharp corners or edges and its dependency on the shape of the crystal, and the strength of the applied shock; 2) debonding between crystal and binder interfaces; 3) collapse of voids in the binder located near an HMX crystal; and 4) the collapse of voids within HMX crystals. Insights are obtained into the relative contributions of these mechanisms to the ignition and growth of hotspots. Understanding these mechanisms of energy localization and their relative importance for hotspot formation and initiation sensitivity of PBXs will aid in the design of energetic material-driven systems with controlled sensitivity, to prevent accidental initiation and ensure reliable performance.

## 1. Introduction

A variety of propulsion and energy delivery systems contain plastic bonded explosives (PBXs), which are heterogenous mixtures of organic (typically CHNO) crystals of energetic materials such as HMX, RDX, TATB etc., and polymeric binders [1] along with other additives. For safety and reliability of systems equipped with heterogeneous energetic (HE) materials, it is desired that PBXs are insensitive to initiation under accidental mechanical insults while retaining designed performance under normal operating conditions. It is well recognized that the shock-induced initiation of PBXs occurs through the formation of hotspots [2–5], which are sites of energy localization resulting from the interaction of an incident shock with the microstructure of the heterogenous composite energetic material. Hotspots with sufficient thermal energy content will initiate self-sustained exothermic chemical reactions [6]. These so-called "critical" hot spots can grow to occupy the volume of the PBX material, with the reactions progressing through deflagration or detonation, depending on the rate of heat release and its coupling with the passing shock wave [7–9]. Understanding the mechanisms leading to shock-induced hotspot initiation is essential for designing PBXs with desired shock sensitivity and predicting their response to accidental mechanical insults. In this work, the mechanics of shock interaction with isolated crystals of HMX embedded in an Estane binder is studied through numerical simulations, to reveal the effects of microstructural defects, such as crystal surface features, and intra- and extra-crystalline pores on the ignition characteristics of PBXs under shock loading.

Irregularities such as crystal-binder interfaces with sharp corners, debonded regions between the crystal and binder, as well as pores in the binder and crystal phases, are observed to be abundant in microstructural images of PBXs [10–13]. PBX samples containing crystals of EMs with sharp corners or rough surfaces have been shown to exhibit higher sensitivity to shocks [14]. The presence of inter- and intra- granular defects in PBXs is also well known to increase the shock-sensitivity of the PBXs [15–17]. While there is strong, albeit circumstantial, evidence of the effects of different types of heterogeneities on shock sensitivity, it is difficult to directly observe the mechanisms of hotspot formation at sites of microstructural heterogeneities, due to the small length and time scales involved as well as due to the lack of optical access and resolution to observe the evolution of the hotspots. To date, the formation and growth of hotspots in PBXs have only been observed directly in a few experimental studies [10, 18–20]. In their recent work, Johnson et al. [10] presented important insights into the hotspot formation process at the microscale by spatio-temporally resolving shock interactions with a single crystal embedded in a binder. These experiments suggest that shock-induced hotspots may appear at crystal-binder interfaces and crystal-crystal junctions in addition to the well-known mechanism of hotspot formation due to the collapse of internal voids within the EM crystal. Johnson et al. [10] observed hotspots forming preferentially on corners or edges of crystals. The maximum temperature at these hotspots was estimated to be about 4000 K. For cases with multiple hotspots, the resulting flame front was seen to propagate along crystal edges, with crystals of about 200 µm size mostly combusted after about 300 ns. In polycrystalline grains, even higher temperature (~6000 K) hot spots were created near internal defects or crystal junctions. However, the thermal mass of the material at these high temperatures of ~6000 K was quite small so that after these hotspots cooled down, HMX combustion proceeded in a similar manner to those of single crystals. Based on their experiments with several crystals, it was speculated in [10] that shock focusing at edges and corners or collapse of the intra- and inter- granular voids are all mechanisms that can lead to formation of hotspots. However, the mechanistic details of how the interaction between an applied shock and the various types of microstructural irregularities in PBXs lead to the formation of hot spots at different types of localization sites are not yet well understood. For example, Johnson et al. [10] indicate that the formation of hotspots at edges and corners of the crystal may be due to shock focusing or due to possible delamination of the binder at the sharp corners. Whether one or more of these mechanisms can lead to hotspot initiation at the surface features (asperities, delaminated zones) on the crystals is not clear. In addition, while there is a difference in temperature between the surface hotspots and the internal hotspots observed in their experiments, with the latter reported to be hotter than the former, the relationship between the observed differences in sensitivity and the (possibly) different mechanisms of hotspot formation remain to be elucidated.

Numerical studies provide an alternative route to understanding the mechanisms responsible for shock-induced energy localization in PBXs. Most previous studies of ignition mechanisms of PBXs have focused on the dissipative mechanisms of hotspot formation, e.g., due to friction and visco-plastic work [21–23]. Studies of hotspots generated by pore/void collapse are abundant [24–28], but typically assume that the voids are located in a background material that is modeled as the homogeneous energetic crystal (e.g., HMX [29]). Hotspot formation due to friction [21, 30] and void collapse [31] in real/realistic microstructures have also been studied, but such studies have inadequately resolved interfaces and the interaction between the crystals, voids, and binder phases. While pore collapse within energetic crystals and resulting hotspot evolution are well represented by many numerical studies [27, 29, 31–36], the presence of pores in the binder or near the crystal-binder surface may also lead to hotspots. In recent work, Springer et al. [27] numerically studied the mechanism of shock-induced critical hotspot formation in HMX grains, with internal and surface pores of analytical shapes embedded in a Kel-F binder. They showed that the location of the pore, as well as the shock strength, has a significant effect on the hotspot formation. In general, a pore in the binder or at the debonded zone between the binder and a crystal may also collapse and, akin to hydrodynamic cavitation bubble collapsing near a solid surface [27], lead to the formation of intense hotspots.

Shock-induced energy localization in rough/faceted energetic crystals of realistic shapes in PBXs has also not been thoroughly studied through interface-resolved simulations. MD calculations performed by An et al. [37, 38] showed the possibility of hotspot formation at sharp corners of crystals in PBXs. The propensity for hotspot formation was directly linked to shock focusing in their work, with concave (towards the shock) corners presenting sites of focusing and hotspot generation. The direction of shock arrival, i.e., from the binder into the crystal or vice-versa, is also seen to determine whether focusing occurs. This shock focusing effect at concave corners was also shown in water-solid systems in experiments conducted by Wang and Eliasson [39]. Grain-scale continuum calculations of Baer [16] showed that shock interaction with microstructural heterogeneities such as extra-granular voids and interactions between crystals in PBXs may lead to formation of hotspots. However, previous calculations of microstructures were not well enough resolved [16] to elucidate the role played by interfaces between crystals or between crystal and binder in energy localization. We show in this paper that rather stringent resolution requirements are necessary to capture all the possible mechanisms for energy localization in PBXs with real crystal geometries that present asperities, corners, facets, and imperfections both within the crystal and at the crystal-binder interfaces.

Various mechanisms of hotspot formation in PBXs due to shock interaction with microstructural heterogeneities are studied in this work through interface-resolved reactive numerical calculations at the grain scale. We show that high levels of resolution, afforded by dynamic solution-adaptive meshing is required to capture shock focusing and initiation of reactions at interfaces. The need for high resolution, even in calculations involving a single pore in a uniform HMX matrix, has been pointed to in previous work by Rai *et al*.[34]; inadequate resolution can not only smear out the physics of the interaction of shocks with interfaces, but the subsequent evolution of the hotspot may also be incorrectly represented, leading to quenching of weak hotspots due to (physical) thermal diffusion or enhancement of growth rates due to numerical diffusion. Here, to maintain adequate resolution throughout the process of shock interactions with isolated HMX crystals of realistic shapes embedded in an Estane binder, adaptive local mesh refinement (LMR) is employed along with sharp interface tracking in a Cartesian grid-based Eulerian calculation [40–42]. In the present framework, the shape of a realistic HMX crystal can be directly imported into the computational domain from micro-CT images [43]. Some of the frequently observed microstructural features in the CT images are crystals with sharp edges, debonded regions between binder and crystal interfaces, voids in binder and within the crystal. By systematically investigating how these microstructural features contribute to energy localization and hotspot formation in PBXs under shock loading, the present paper elucidates various possible modalities for hotspot initiation in realistic (i.e., rough and faceted) crystals placed in a binder. Note however that the calculations presented are in 2D, for which the high resolution employed already demands intensive computational calculations. Performing highly resolved 3D calculations for real crystals is certainly required but would be computationally extremely expensive and is therefore deferred to future work.

The remainder of the paper is organized as follows: numerical methods employed for the interface-resolved reactive simulations of shock interaction with HMX crystals are described in Section 2. Shock-induced energy localization at the sharp corners of crystal-binder interfaces is investigated in Section 3.1. The contribution of delaminated regions at the crystal-binder interface to critical hotspot formation in HMX crystals under shock loading is studied in Section 3.2. Energy localization in the HMX crystals due to the shock-induced collapse of a void in the binder phase is studied in section 3.3. Shock interaction with a HMX crystal of realistic shape in an Estane binder, obtained from micro-CT imaging, is studied in Section 3.4, which combines the different modalities of hotspot formation, i.e., from surface asperities, debonded zones and voids in the crystal. Conclusions from the study are summarized in Section 4.

## 2. Methods

The equations governing motion and deformation of the crystal and binder materials are cast in Eulerian form and numerically solved using SCIMITAR3D, a Cartesian grid-based sharp-interface framework for compressible multi-material flows [40, 42]. The interfaces between the crystal, binder, and void are embedded in the Cartesian grid and tracked using the levelset method [44]. The ghost fluid method (GFM) is used to impose appropriate boundary conditions at the material-material and material-void interface [40]. The numerical framework used in this work has been described in detail in previous works [33, 36, 40, 41, 45, 46] and validated against experiments[36, 40, 47] and molecular dynamics simulations [35] for high-speed multi-material shock and impact problems. The governing equations and material models for HMX and Estane are briefly described in the following subsection.

## 2.1 Conservation laws

The mass, momentum, and energy conservation equations cast in the following form are solved independently for each material (crystal and binder):

$$\frac{\partial \rho}{\partial t} + \frac{\partial (\rho u_i)}{\partial x_i} = 0 \qquad (1)$$

$$\frac{\partial (\rho u_i)}{\partial t} + \frac{\partial (\rho u_i u_j - \sigma_{ij})}{\partial x_j} = 0 \qquad (2)$$

and

$$\frac{\partial (\rho E)}{\partial t} + \frac{\partial (\rho E u_j - \sigma_{ij} u_i)}{\partial x_j} = 0 \qquad (3)$$

where $\rho$, and $u_i$, respectively denote density, and velocity components, $E = e + \frac{1}{2} u_i u_i$ is the specific total energy, and $e$ is the specific internal energy. $\sigma_{ij}$ is the Cauchy stress tensor:

$$\sigma_{ij} = S_{ij} - p\delta_{ij} \qquad (4)$$

where $S_{ij}$ is the deviatoric stress tensor and $p$ is the pressure. $p$ is obtained from an equation of state (EOS), as discussed separately for crystal and binder materials below. The deviatoric stress components, $S_{ij}$ are evolved using the following equations,

$$\frac{\partial (\rho S_{ij})}{\partial t} + \frac{\partial (\rho S_{ij} u_k)}{\partial x_k} + \rho S_{ik} \Omega_{kj} - \rho \Omega_{ik} S_{kj} = 2\rho G D_{ij}^{\mathrm{d}} \qquad (5)$$

where $D_{ij}^{\mathrm{d}}$ is the deviatoric component of the strain rate tensor, $\Omega_{ij}$ is the spin tensor, and $G$ is the shear modulus of the material.

The chemical decomposition of HMX is modeled using a global 3-step mechanism involving four different groups of species proposed by Tarver *et al.*[6]. The mass-fraction of the chemical components/species in this rection model is tracked by solving the species conservation equation:

$$\frac{\partial \rho Y_\mathrm{i}}{\partial t} + \frac{\partial}{\partial x_j} (\rho Y_\mathrm{i} u_j) = \dot{Y}_\mathrm{i} \qquad (6)$$

where $Y_i$ is the mass-fraction of $i^{\mathrm{th}}$ species and $\dot{Y}_\mathrm{i}$ represents the rate of production or destruction of the $i^{\mathrm{th}}$ species through chemical reactions. The implementation of the 3-step model in the current numerical framework is presented in previous work [31, 33, 34, 48] and given in the Appendix for completeness.

The change in temperature due to chemical decomposition of HMX is calculated by solving the evolution equation for temperature($T$),

$$\rho c_p \dot{T} = \dot{Q}_R + k\nabla^2 T \tag{7}$$

where $c_p$ is the specific heat of the reaction mixture at constant pressure, $\dot{Q}_R$ is the total heat release rate because of the chemical reaction, and $k$ is the thermal conductivity of the reaction mixture. The calculation of $c_p$, $\dot{Q}_R$, and $k$ was discussed in previous work [34] and given in the Appendix for completeness.

The solid/liquid phase of HMX during the calculation is decided based on the melting point of HMX ($T_m$) computed as follows[49]:

$$T_m = T_{m0}\left(1 + \frac{p - p_{\text{ref}}}{a'}\right)^{1/c'} \tag{8}$$

where the reference pressure $p_{\text{ref}} = 0$ and the reference melting temperature $T_{m0} = 551$ K. $a'$ and $c'$ are fitting parameters and their values are 0.305 GPa and 3.27 respectively[49].

## 2.2 Material model for HMX

Below its melting point, HMX is modelled as a solid elasto-plastic material. In this case, Eq. (5) is solved using a two-step operator splitting algorithm[50]. First, the deviatoric stress is evolved assuming elastic deformation in a predictor step. This is followed by a correction step to remap the predicted stress onto a yield surface using the radial return algorithm [51]. The J2-plasticity theory is used, with the consistency condition:

$$f = S_e - \sigma_Y = 0 \tag{9}$$

where $S_e\left(=\sqrt{\frac{3}{2}tr(S_{ij}S_{ji})}\right)$ is the von Mises stress and $\sigma_Y$ is the yield strength of the material. For the elastic-perfectly-plastic model $\sigma_Y$ is set to a constant value of 260 MPa and $G = 10$ GPa [52].

The HMX in the computational domain above the melting point (given by Eq. (8)) is modelled as a viscous liquid. The deviatoric stresses in the melted HMX are computed from:

$$S_{ij} = 2\mu D_{ij}^d \tag{10}$$

where $\mu$ is the shear viscosity of the melted HMX and is computed from the temperature and pressure dependent relation given by Kroonblawd and Austin[49].

The Mie-Grüneisen equation of state (EOS) for HMX of the following form is used to obtain $p$ [53]:

$$p = p_c(V) + \frac{\Gamma}{V}[e - e_c(V)] \tag{11}$$

where $\Gamma$ is the Grüneisen coefficient and $V$ is the specific volume. The pressure $p_c$ on the cold curve is fit to a Birch-Murnaghan EOS[54]:

$$p_c = \frac{3}{2} K_0 \left[ \left(\frac{V}{V_0}\right)^{-\frac{7}{3}} - \left(\frac{V}{V_0}\right)^{-\frac{5}{3}} \right] \left[ 1 + \frac{3}{4}(K_0' - 4) \left\{ \left(\frac{V}{V_0}\right)^{-\frac{2}{3}} - 1 \right\} \right] \tag{12}$$

where the coefficients $K_0$ and $K_0'$ are the calibrated isothermal bulk modulus and its pressure derivative [55, 56], respectively, and $V_0$ is a reference specific volume; values of these constants are provided in Table 1.

| Parameter | Value |
| --- | --- |
| $V_0$ | $\frac{1}{1900}$ m³/kg |
| $K_0$ | 16.5 GPa |
| $K_0'$ | 8.7 |
| $\Gamma_0$ | 1.3 |
| $Z$ | 0.2 |

Table 1 The EOS parameters for HMX used in the continuum calculations.

The internal energy corresponding to the isothermal state is obtained from

$$e_c(V) = e_0 - \int_{V_0}^{V} p_c(V) dV \tag{13}$$

where $e_0$ is a reference internal energy. The Grüneisen coefficient is defined as:

$$\Gamma(V) = a + b \left(\frac{V}{V_0}\right) \tag{14}$$

and the coefficients $a$ and $b$ are calculated from the calibration factor $Z$ [57] and zero-pressure Grüneisen parameter $\Gamma_0$ [58]:

$$a = \Gamma_0 + b \tag{15}$$

$$b = -Z \tag{16}$$

The EOS parameters $V_0, K_0, K_0', Z,$ and $\Gamma_o$ are given in Table 1.

### 2.3 Material model for binder

The Estane binder is modelled as an inert visco-elastic material. The shear modulus $(G)$ in Eq. (5) is computed from a Prony series representation of the generalized Maxwell model given below:

$$G(t) = G_0 + \sum_{i=1}^{n} G_i \exp\left(-\frac{t}{\tau_i}\right) \tag{17}$$

where $n = 22$, $G_0$ is the shear modulus of binder under zero strain. The fitting parameters $G_i$ and $\tau_i$ are taken from [59].

The Mie-Grüneisen equation of state (EOS) in the following form is used to compute pressure in binder from the specific internal energy:

$$p(e, V) = \Gamma \frac{e}{V} + \begin{cases} \frac{K_0 \vartheta}{(1 - s\vartheta)^2}\left[1 - \frac{\Gamma}{2V}(V - V_0)\right], & if\ V \leq V_0 \\ K_0\left(\frac{V_0}{V} - 1\right), & V > V_0 \end{cases} \tag{18}$$

Here $\vartheta$, $K_0$ and $\Gamma_0$ are the volumetric strain, bulk-modulus, and Grüneisen parameter respectively. They are computed as follows:

$$\vartheta = 1 - \frac{\rho_0}{\rho} \tag{19}$$

$$K_0 = \rho_0 c_0^2 \tag{20}$$

$$\Gamma(V) = \frac{\Gamma_0 \rho_0}{\rho} \tag{21}$$

In the above equation, $\rho_0, c_0,$ and $\Gamma_0$ are the reference density, reference speed of sound and the reference Grüneisen parameter. These material parameters for the binder are given in Table 2 [47].

| Parameter | Value |
|---|---|
| $\rho_0$ | 1190 kg/m$^3$ |
| $c_0$ | 1750 m/s |
| $\Gamma_0$ | 1 |
| $s$ | 2.2 |

Table 2. EOS and material parameters of the Estane binder used in this work.

The temperature in the binder is computed from the specific internal energy:

$$T = T_0 + \frac{(e - e_0)}{c_v} \tag{22}$$

where the reference temperature in binder, $T_0$ is 298.14 K and the reference specific internal energy of binder, and $e_0 = 0$ J/kg, $c_v = 2076$ J $\cdot$ kg$^{-1}$ $\cdot$ K$^{-1}$[47].

### 2.4 Defining and tracking crystal-binder, crystal-void and binder-void interfaces using levelsets

Narrow-band levelset tracking[44] is used in the current framework to sharply define and evolve all material interfaces. The levelset field is advected to capture the evolution of the interface via:

$$\frac{\partial \phi}{\partial t} + \boldsymbol{u_n} \cdot \nabla \phi = 0 \tag{23}$$

where $\phi$ represents the levelset field and $\boldsymbol{u_n}$ is the velocity of the corresponding material interface. Details of the implementation of narrowband levelset tracking in the current framework can be found in [40–42, 50].

For cases where the simulations are conducted on real, i.e., imaged crystal-binder geometries, levelset fields must be created in the computational domain by importing and processing images, which reside in the voxelized image domain. The initial levelset field in the computational domain representing the shape of the crystal and binder is obtained from grayscale images (see Figure 1) using an image processing framework [43]. First, a speckle reducing anisotropic diffusion method [60] is used to remove noise from the image intensity field. The crystal shape is then extracted from the denoised image intensity field using an active contour method [61]. The levelset fields delineating the shape of the crystal-binder interface in the computational domain are obtained from this active contouring algorithm. The levelsets are then embedded in a uniform Cartesian grid on which the flow computations are performed. A local mesh refinement (LMR) algorithm [62] then refines the mesh in the vicinity of interfaces, and as the flow evolves, also adaptively resolves shocks and other gradients in the flow.

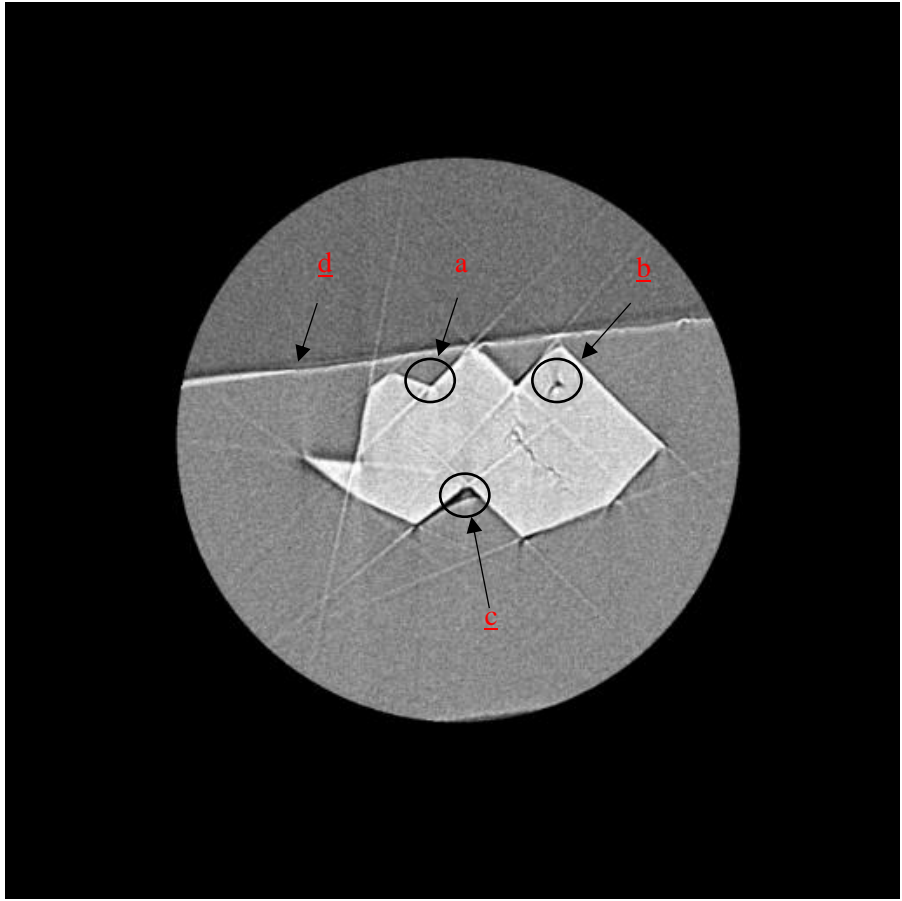

Figure 1. CT scan of HMX crystal in binder. Noticeable features of the HMX crystal: (a) sharp corner at crystal-binder interface, (b) void in HMX crystal, (c) debonding between the crystal and binder at the interface, and (d) binder air interface.

In this work, the levelset approach for tracking isolated interfaces is modified to accommodate the crystal-binder interface, which involves two coincident interfaces that can locally separate by debonding, viz., the crystal-void and binder-void interface. For this purpose, separate levelset fields are used to describe the crystal-void and binder-void interface. When the binder and crystal are in cohesion, the crystal-void and binder-void interfaces are exactly coincident, with no intervening void space. When the binder and crystal are not in perfect contact a void space may be formed in the debonded region. By convention [44], a levelset field representing a specific (single) material carries a positive sign within that material and a negative sign outside the material; e.g., in the configuration shown in Figure 2 where an HMX crystal of circular shape (orange) is embedded in the binder (blue). A levelset field $\phi_{\text{HMX}}$(Figure 2 (a)) is used to describe the HMX crystal such that $\phi_{\text{HMX}} > 0$ inside the HMX crystal and $\phi_{\text{HMX}} < 0$ outside the HMX crystal. A separate levelset field $\phi_{\text{binder}}$(Figure 2 (b)) is used such that $\phi_{\text{binder}} > 0$ in the binder and $\phi_{\text{binder}} < 0$ outside the binder. The zero levelset contours of $\phi_{\text{crystal}}$ and $\phi_{\text{binder}}$ coincide at the interface of crystal and binder, except at the delaminated region in Figure 2 (c). Therefore, $\phi_{\text{binder}}$ and $\phi_{\text{HMX}}$ are both negative in the white region at the center of the HMX crystal and in the delaminated region in Figure 2 (c).

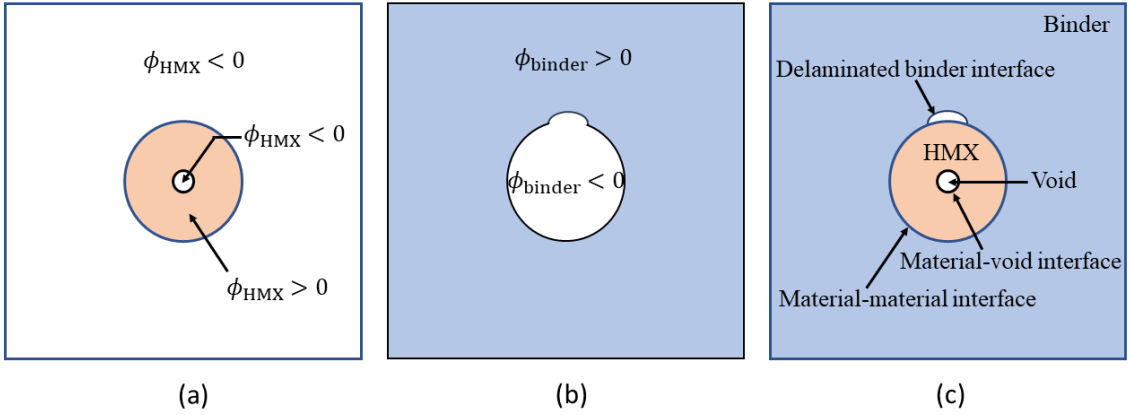

Figure 2. A schematic diagram of a circular HMX crystal with a void at its center embedded in binder to demonstrate the sign convention used to define separate levelset fields describing the crystal and the binder region. (a) shows the levelset field for HMX crystal, (b) shows the levelset field for binder, and (c) shows the crystal-binder interface with a delaminated region.

Appropriate boundary conditions are applied at the material-material (crystal-binder) and the material-void (crystal-void, binder-void) interfaces [40]. Frictionless contact is assumed at the material-material interfaces, i.e., continuity of the normal velocity and normal traction in the two materials is imposed [47]. The tangential components of velocity and traction in the two materials are assumed to be discontinuous at the interface to allow for frictionless sliding[47]. A free surface condition, as described in [40], is used at the material-void interfaces. These boundary conditions at the material-material and material-void interfaces are applied using the ghost fluid method [63]. The interfacial conditions and their implementation in the current framework are described in detail in previous works [40, 47] and are not repeated here in the interest of brevity.

### 2.5 Numerical solution techniques for solving the governing equations

Numerical schemes and algorithms for solving the system of equations in the crystal and binder phases have been described in detail in previous works [40, 42, 47]. Briefly, the convective terms on the right-hand side of the governing equations (1) – (3) and (5) are spatially discretized using a third-order essentially non-oscillatory scheme (ENO)[64] and the time-integration of these equations is performed using a third-order total variation diminishing (TVD) Runge-Kutta scheme [65]. The species evolution equations, Eq. (6), are integrated in time using the Strang operator splitting approach described in [31, 48] to circumvent numerical stiffness due to the reactive source terms. A fifth-order Runge-Kutta Fehlberg method [33] method is used to integrate the stiff reactive source terms in Eq. (6) and (7). Further details on the numerical implementation of the solver in the levelset based interface capturing framework can be found in [42, 50, 66]. The current numerical framework has been validated against experimental results for shock induced collapse of a spherical void in PMMA[36], molecular dynamics based calculations of void collapse in TATB [32] and HMX [35], and benchmark experimental and numerical results for high speed shock and impact problems [40, 47]. An Octree-based adaptive local mesh refinement (LMR) [62] approach is used to provide high resolution of all interfaces, shocks, and reaction fronts. Four levels of refinement are employed in this work, with the finest meshes adapted to resolve high gradients of pressure, temperature and the levelset narrowband as the solution develops. Further details of the LMR implementation for shock and interface dominated problems can be found in previous work [62].

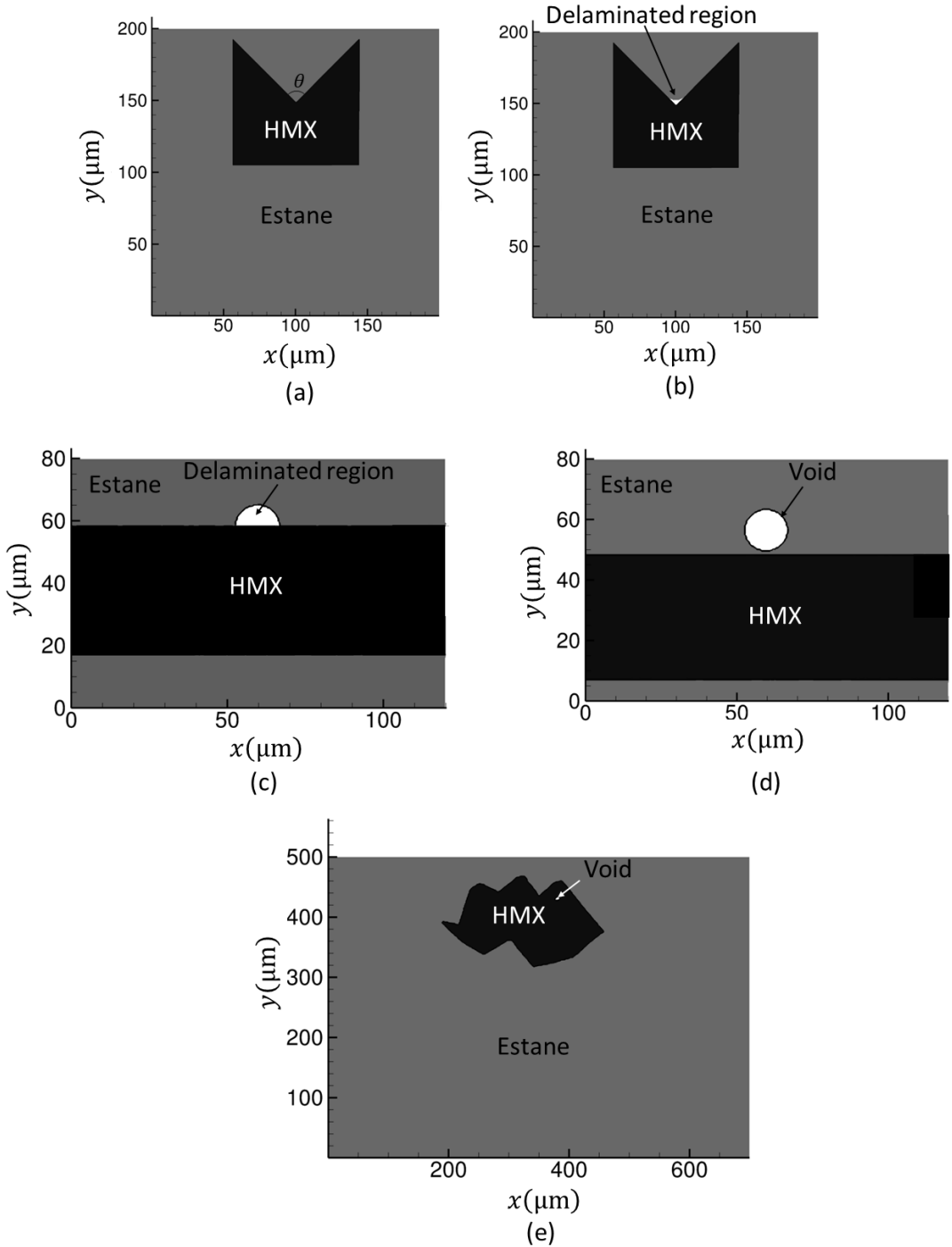

Figure 3. Initial setup of the calculation of shock-interaction with a) a synthetic HMX crystal embedded in Estane, b) a small delaminated region of 9 μm radius of curvature located at the concave corner, c) a semi-circular delamination void at a planar crystal-binder interface, d) an HMX crystal embedded in Estane binder with a void (14 μm in diameter, centered at 15.4 μm away from the interface) in the binder located near the crystal-binder interface, and e) HMX crystal with defects.

## 2.6 Grid-independence study

The numerical framework described above was validated against experiments[36, 40, 47] and molecular dynamics simulations [35] for high-speed multi-material shock and impact problems. In this sub-section, a grid-independence study is carried out to establish the grid resolution required for the calculations of shock interaction with HMX crystal embedded in Estane binder. Shock interaction with an HMX crystal of idealized shape embedded in Estane binder is selected for the grid independence study. Figure 3 (a) shows a schematic of the computational setup for this calculation. The exterior angle between the two faces of the crystal (in Figure 3 (a)) at the concave corner, $\theta$, is selected as the control parameter representing the shape feature of the concave corner. A supported shock load is applied at the top boundary using a constant particle velocity $(U_p)$. This boundary condition is implemented by setting $v = -U_p$ at the top boundary of the domain, while a Neumann boundary condition is used for $\rho, u, p,$ and $S_{ij}$. The east, west, and south boundaries of the domain are treated as outflow boundaries, i.e., Neumann boundary conditions for $\rho, u, v, p,$ and $S_{ij}$ are applied at these boundaries as well. Unless otherwise specified, these boundary conditions are used in all the calculations presented in this paper.

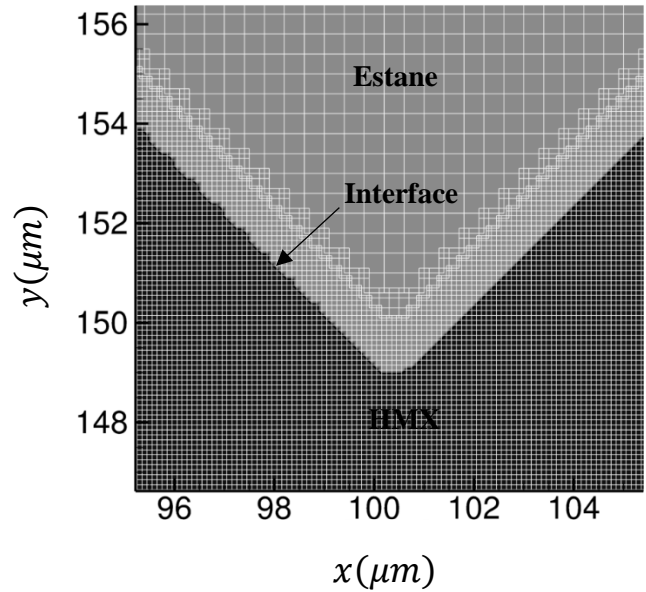

Figure 4. Locally refined mesh around the interface of HMX crystal and Estane binder.

A tree-based local mesh refinement strategy is employed to reduce the computational cost of modelling the interaction of shock waves with interfaces. Initially, the entire computational domain is resolved with a uniform Cartesian mesh of size $\Delta x_{\text{base}}$ (given in Table 3). In the current calculations, the HMX crystal and region near the crystal-binder interface are resolved at the mesh level $n = 3$. At the $n^{\text{th}}$ level of refinement$(n \in N)$, the mesh size is given by:

$$\Delta x_n = \frac{\Delta x_{\text{base}}}{2^{n-1}} \tag{24}$$

The locally refined mesh within the crystal and around the interface is demonstrated in Figure 4. The rest of the computational domain (i.e. the binder) is selectively refined with $n = 3$ around the compression and rarefaction waves, based on the density gradients in the field[62]. For this grid-independence study, the response of a HMX crystal with $\theta = \frac{\pi}{2}$ under the action of a sustained shock of $U_p = 2500$ m/s is computed using different grid resolutions. Four different grid resolutions used are tabulated in Table 3.

|  | $\Delta x_{base}$ | $n$ | $\Delta x_3$ | $\epsilon_{GRID\ i}$ |
| --- | --- | --- | --- | --- |
| GRID 1 | 1.6 μm | 3 | 0.4 μm | 0.208 |
| GRID 2 | 0.8 μm | 3 | 0.2 μm | 0.11 |
| GRID 3 | 0.4 μm | 3 | 0.1 μm | 0.003 |
| GRID 4 | 0.2 μm | 3 | 0.05 μm | |

Table 3. Grid resolutions used for the calculation of $U_p = 2500$ m/s shock interaction with a representative HMX crystal with $\theta = \frac{\pi}{2}$

The total mass of gaseous reaction products ($F$) in the 2D plane of calculation is chosen as the quantity of interest (QoI) to measure ignition and reaction progress in the crystal. This QoI is calculated as:

$$F = \int_A \rho Y_4 dA \tag{25}$$

where $Y_4$ is mass fraction of the gaseous products of combustion of HMX in the reduced-order 3-step reaction mechanism of HMX given by Tarver et al.[6] (see Appendix for the reaction chemistry model). Note that since $F$ is computed by integrating $\rho Y_4$ over a 2D domain, we obtain this QoI in units of kg/m. Figure 5 compares the values of $F$ obtained from the calculations using the four different grid resolutions in Table 3, showing that the results converge to the solution obtained from GRID 4. The relative error ($\epsilon_{GRID\ i}$) in the prediction of $F$ obtained from GRID $i$: $i \in \{1, 2, 3\}$ with respect to GRID 4 is:

$$\epsilon_{GRID\ i} = \left| \frac{F_{GRID\ i} - F_{GRID\ 4}}{F_{GRID\ 4}} \right| \tag{26}$$

where $F_{GRID\ i}$ and $F_{GRID\ 4}$ are the predicted values of $F$ at t=15 ns using intermediate grid sizes GRID $i$ and the finest grid GRID 4, respectively. Table 3 shows that $\epsilon_{GRID\ 3}$ reaches a low value of 0.003 and Figure 5 shows that $F$ computed from GRID 3 is in good agreement with GRID 4. To balance the computational cost with accuracy, GRID 3 is used to obtain grid independent solutions for the rest of the calculations presented in this paper. With this grid arrangement, the salient features (e.g., shocks and multi-material interfaces) in the computation domain are adequately resolved.

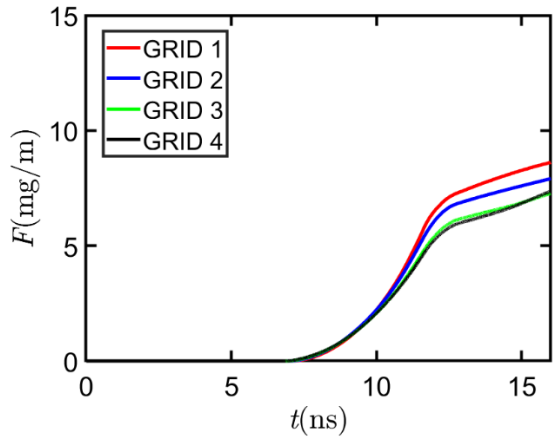

Figure 5. A grid sensitivity study: comparison of F computed using different grid resolutions ($\Delta x_{\text{base}} = 1.6\ \mu m, 0.8\ \mu m, 0.4\ \mu m$, and $0.2\ \mu m$)

## 3. Results and discussion

In this section, the numerical framework described above is employed to investigate the mechanistic details of energy localization due to shock interactions with the microstructural heterogeneities in PBXs, resolving phenomena at the length scale of individual, embedded HMX crystals. The important microstructural features can be observed in the 2D micro-CT cross-section of a single HMX crystal shown in Figure 1. First, in Section 3.1, we examine the possibility of critical hotspot formation at the sharp corners (feature *a* in Figure 1) of crystal-binder interface under shock loading. Then, in Section 3.2, we examine the effects of energy-localization in PBXs due to shock interaction with debonded regions (feature *c* in Figure 1) which form voids at the crystal-binder interface. Ignition of PBXs due to shock-induced collapse of a void in binder near a HMX crystal is studied in section 3.3. Finally, in section 3.4, we study the effect of defects (voids and cracks, feature *b* in Figure 1) within the crystal (as shown in Figure 1) on its ignition and combustion under shock loading.

### 3.1 Effect of sharp corners on HMX crystals in a binder

Energetic crystals in PBXs often have sharp corners[14]; e.g., the micro-CT imaged cross-section of the HMX crystal in Figure 1 shows sharp convex as well as concave corners. Preliminary studies showed that convex corners were less prone to forming hotspots than concave ones. This is in agreement with previous studies [37, 38] that showed that energy localizes at corners of concave (facing the shock) aspect for crystals embedded in binders. Similar shock focusing effects have also been seen in solids immersed in liquids [39] and subject to shocks from the liquid side. In this section, we show that such localization of energy at the concave corners of HMX crystals can lead to formation of critical hot spots. The dependence of ignition sensitivity and reaction rate of the HMX crystals on the corner geometry and shocks strength are studied.

To isolate the effects of corners, concave polygon-shaped, internal defect-free, geometries are created to serve as representative (idealized) HMX crystals. Figure 3 (a) shows a schematic of the computational setup for the simulations. The production of gaseous reaction product in an HMX crystal with $\theta = 90^0$ subjected to $U_p = 2500$ m/s shock is shown in Figure 5. The high rate of production of the gaseous reaction products in Figure 5 shows that energy localization due to shock focusing at the concave corner of the HMX crystal leads to self-sustained reaction for $\theta = 90^0$ and $U_p = 2500$ m/s. This shock-focusing induced initiation of the representative HMX crystal for $\theta = 90^0$ is examined through a sequence of temperature contours presented in Figure 6. Figure 6 (a) shows that the temperature within the bulk region of the HMX crystal increases to ~1000 K due to shock compression alone. However, this rise in temperature is not sufficient for ignition within the shock passage timescale. Figure 6 (b) shows that the crystal instead ignites at the sharp concave corner upon the arrival of the shock. Figure 6 (c) shows that after ignition, the reaction front remains coupled with the transmitted shock and travels as a reaction wave within the crystal. Therefore, in this case, a critical hotspot forms due to shock-focusing at the sharp concave corner of the HMX crystal.

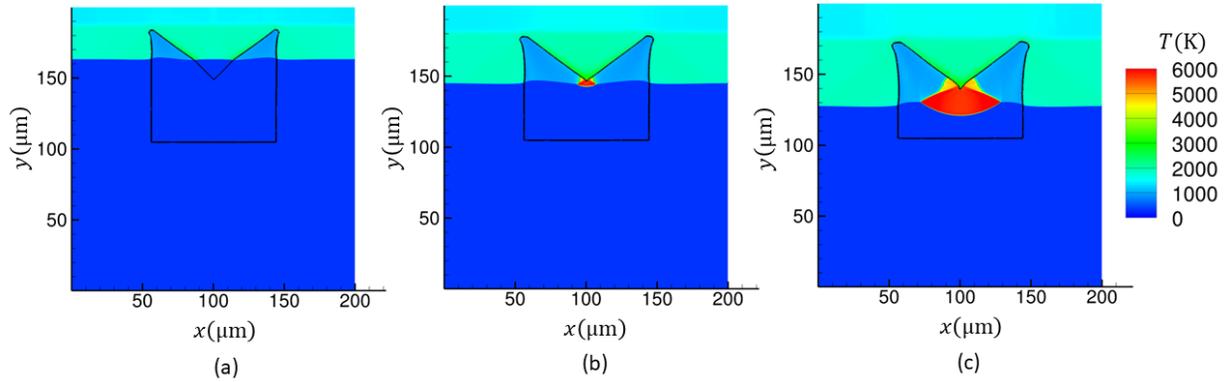

Figure 6. (a) – (c). A sequence of temperature contours obtained from the numerical calculation for $U_p = 2500$ m/s shock interacting with a representative HMX crystal with $\theta = 90^0$ embedded in Estane.

The shock-interface interaction mechanisms leading to initiation at the sharp corner are further investigated by simplifying the above system by reverting to an *inert* calculation, i.e. the reaction terms are switched off in the governing equations. The energy localization due to shock focusing at the sharp concave corner of the crystal can then be demonstrated via the $y - t$ plots in Figure 7. In these plots, pressure$(p)$, the y-component of velocity$(v)$, and temperature$(T)$ along the vertical symmetry line passing through the corner ($i.e.\ at\ x = 100$ μm) are collected over time and arranged chronologically along the $t -$ axis for the duration of the calculation. Figure 7 demonstrates the wave structure that emerges during the interaction of the incident wave with the concave corner. The incident, reflected, and transmitted waves are denoted by I, II, and III in Figure 7, respectively. The strength and speed of the transmitted and the reflected waves depend on the impedance mismatch at the crystal-binder interface and the angle of incidence between the incident shock and the interface. The snapshots of numerical Schlierens in Figure 8 (a) and (b) display the shape of these waves in the computation domain. Figure 8 (a) shows that reflected waves (marked "II") arise due to the angle between the incident wave and the crystal-binder interfaces. These reflected shocks interact as the incident shock travels downwards. The location of these reflected shocks after their interaction is shown in Figure 8 (b). The interaction of the reflected shocks leads to intensification of the $p$, $v$, and $T$ in the binder just above the sharp corner, as seen in Figure 7 (a), (b), and (c) respectively. Therefore, mechanical energy localizes through shock-focusing during the interaction of the incident shock with the corner. The locus of the y-location of the sharp corner at $x = 100$ μm is marked with the black line in Figure 7, showing that the corner itself travels downward following the impingement of the shock. Figure 7 (c) shows that the compression of HMX at the corner increases the temperature at the translating corner to ~1200 $K$ for a sustained period. This sustained localization of thermal energy due to shock focusing at the corner leads to ignition of the HMX crystal in the corresponding reactive calculation, as shown in Figure 6 (b).

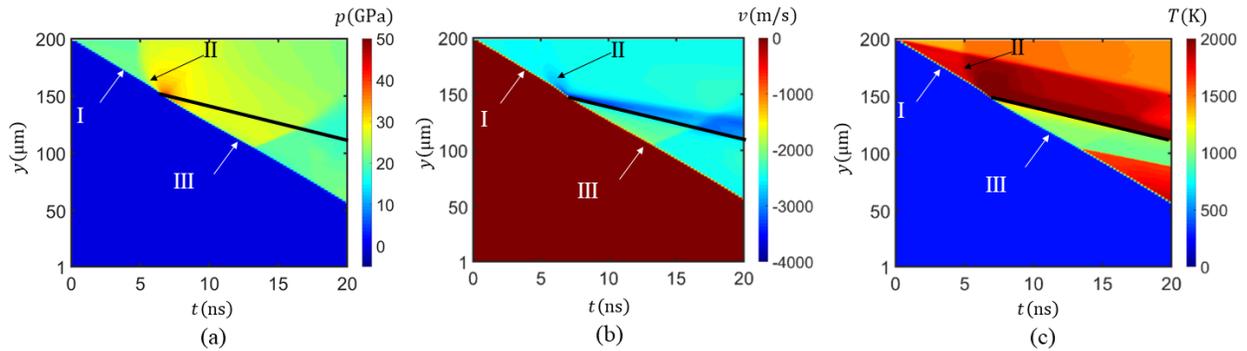

Figure 7. $y - t$ plots of (a) pressure, (b) velocity component along the y-axis and (c) temperature obtained from non-reactive calculation of the $U_p = 2500$ m/s shock interaction with the HMX crystal with $\theta = 90^0$ embedded in binder. The trajectory of the interface is marked with the black line. The markers (I), (II), and (III) show the incident, transmitted, and reflected waves, respectively.

To understand the role of the shape of the corner in the shock response of the representative crystals, three different corner angles $\theta = 60^0$, $90^0$, and $120^0$ are studied, for different shock strengths corresponding to $U_p = 1500$ m/s, $2000$ m/s, and $2500$ m/s. Reactive mechanisms are activated to observe the progress of chemistry following shock interaction with the corners. The comparison of the amount of final combustion product ($F$) for the crystals with $\theta = 60^0$, $90^0$, and $120^0$ and for the $U_p = 1500$ m/s shock in Figure 9 (a) shows that $F \cong 0$ for all three crystals, i.e., none of the crystals ignite at $U_p = 1500$ m/s.

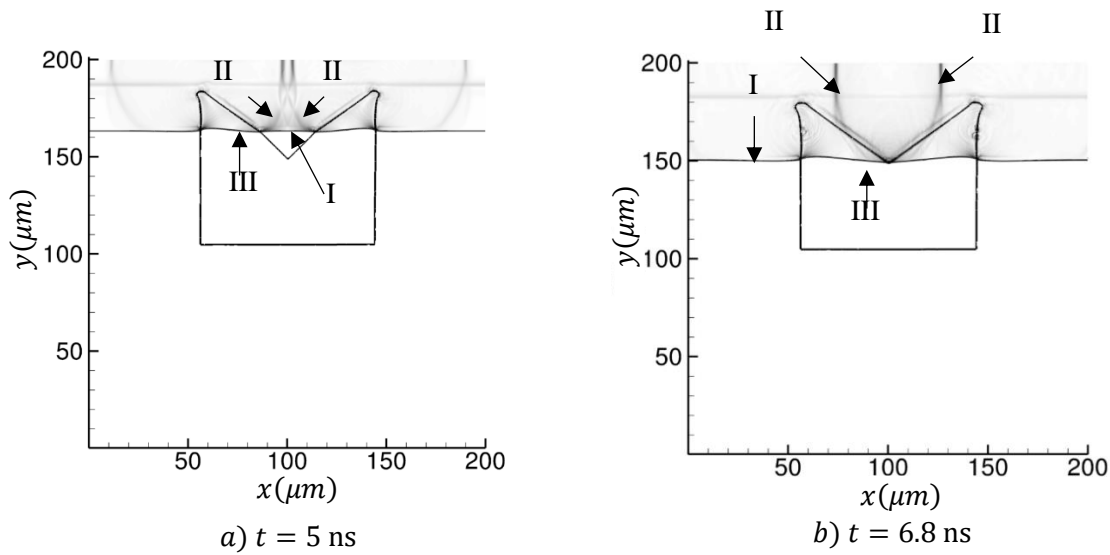

Figure 8. Numerical Schlieren at a) 5 ns and b) 6.8 ns during the $U_p = 2500$ m/s shock interaction with the representative HMX crystal (chemically inert) with $\theta = 90^0$ embedded in Estane binder. The numerical Schlierens show reflected waves (II) generated due to the interaction of the interaction of the incoming shock (I) with the wedge shaped crystal-binder interfaces. The wave transmitted within the crystal is marked with (III).

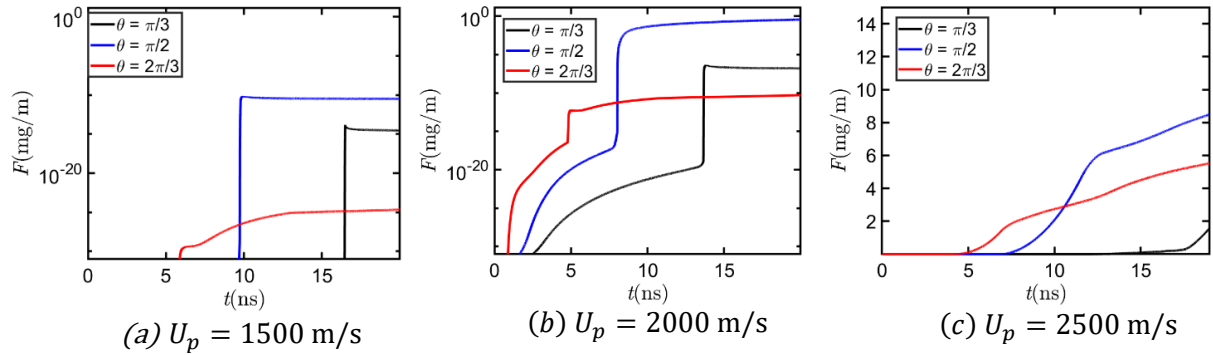

Figure 9. Comparison of the amount of final combustion product($F$) produced in the 2D plane of calculation during shock interaction with synthetic HMX crystals with $\theta = 60^0$, $90^0$, and $120^0$ shown in Figures 3. The response of the three crystals to $U_p = 1500$ m/s, $2000$ m/s, and $2500$ m/s shocks are compared in (a), (b), and (c) respectively.

Figure 9 (b) show that at $U_p = 2000$ m/s only the crystal with $\theta = 90^0$ ignites. Figure 9 (c) shows that all three crystals ignite during interaction with an incoming shock of $U_p = 2500$ m/s. However, the crystal with $\theta = 90^0$ shows a higher reaction rate compared to the other two crystals, i.e., with $\theta = 60^0$ and $120^0$ for the $U_p = 2500$ m/s shock. The strength of the reflected shock shown in Figure 8 and the amount of energy localized in the binder are sensitive to $\theta$ and $U_p$. Consequently, the ignition characteristics of the crystals at the sharp corners due to shock focusing strongly depend on $\theta$ and $U_p$. This study shows that surface characteristics of crystals, such as the angle at sharp concave corners, can play a role in determining the criticality threshold of a particular sample of PBX.

### 3.2 The role of delaminated regions between the binder and crystal in hotspot formation

To examine hotspot formation in PBXs due to shock interaction with a delaminated crystal-binder interface, two different configurations are investigated: delamination voids at sharp corners of the crystal and delamination voids at a planar crystal surface.

3.2.1 Delamination at a sharp corner of HMX crystal

Delamination often occurs due to the lack of perfect cohesion of the binder to the crystal at the sharp concave corners, as seen in Figure 1. Numerical calculations are performed to investigate whether shock-induced collapse of such delaminated regions can augment energy localization at the sharp corners. For this calculation, an HMX crystal with a sharp concave corner of angle $\theta = 90^0$ shown in Figure 3 (b) is considered, with a void space due to delamination of the binder at the corner. To create the void space, the HMX crystal is given a sharp concave corner while the binder interface at the corner has a radius of curvature of 9 μm. The initial shape of the delaminated region between the crystal and binder at the sharp corner is shown in Figure 3 (b), leading effectively to a triangular-shaped void space. Calculations are performed for $U_p = 1500$ m/s, $2000$ m/s, and $2500$ m/s to characterize the sensitivity of delaminated crystal-binder interfaces during shock initiation of the PBXs.

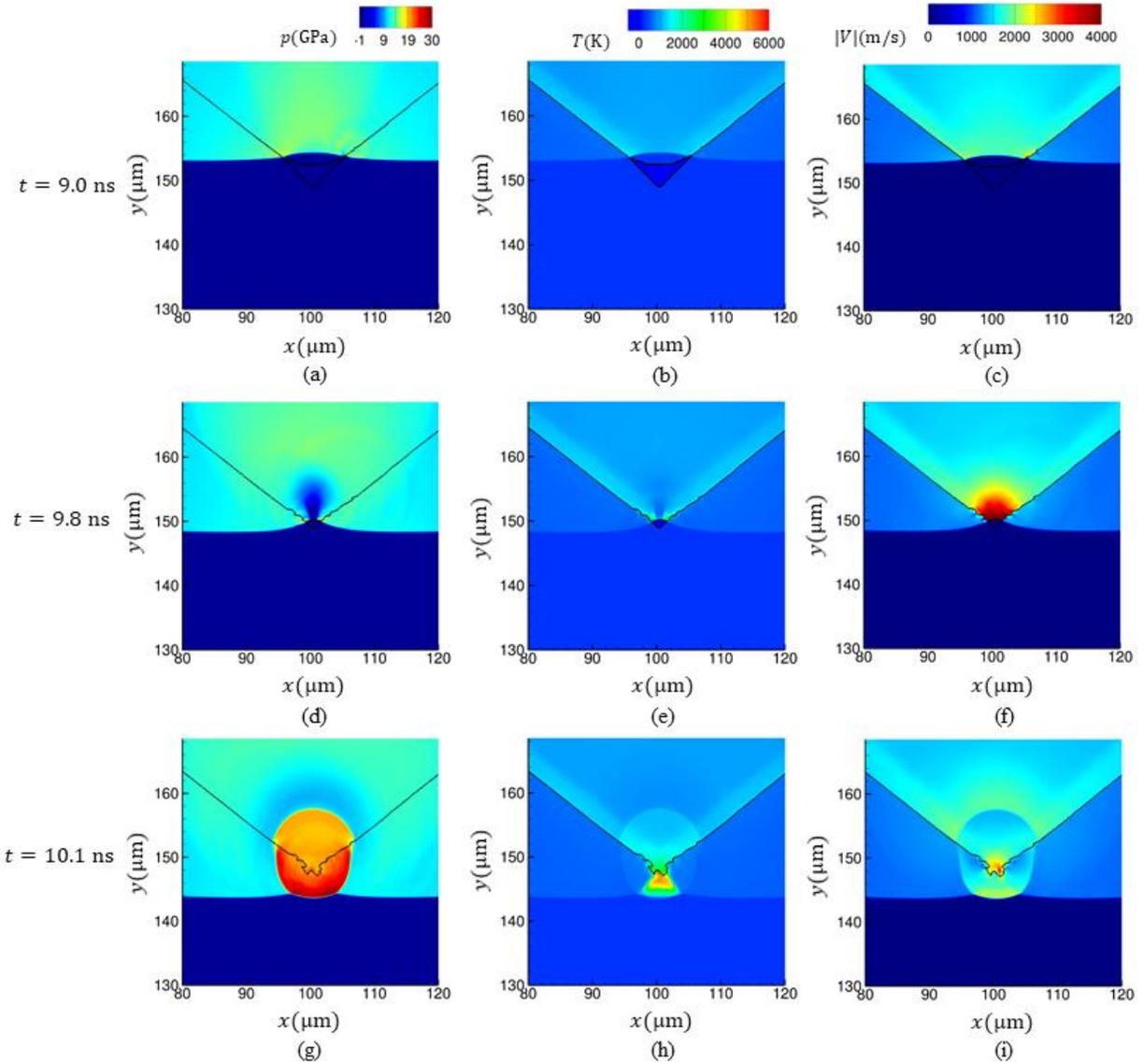

Figure 10. (a) – (i). A sequence of temperature contours obtained from the numerical calculation for $U_p = 1500$ m/s shock, $\theta = 90^0$ and with a delaminated region of 9 µm radius of curvature at the corner. The rows from top to bottom correspond to the times 9.0 ns, 9.8 ns, and 10.1 ns respectively. The columns from left to right show the contours of pressure, temperature, and velocity magnitude, respectively.

Figure 10 shows a sequence of contours of pressure($p$), temperature($T$) and velocity magnitude $\left(|V| = \sqrt{u^2 + v^2}\right)$ in the region around the sharp corner. The results show that a hotspot is formed during the interaction with the incoming $U_p = 1500$ m/s shock. In contrast, the same crystal without a delaminated region at the sharp corner ($\theta = 90^0$) does not ignite when subjected to a $U_p = 1500$ m/s shock (Figure 9 (a)). The contour plots in the top row of Figure 10 are at the instant ($t = 9$ ns) when the incident shock arrives at the void. The pressure and temperature contours in Figure 10 (d) and (e), respectively, show that the incident shock reflects as a rarefaction wave from the void surface. The binder-void interface is then

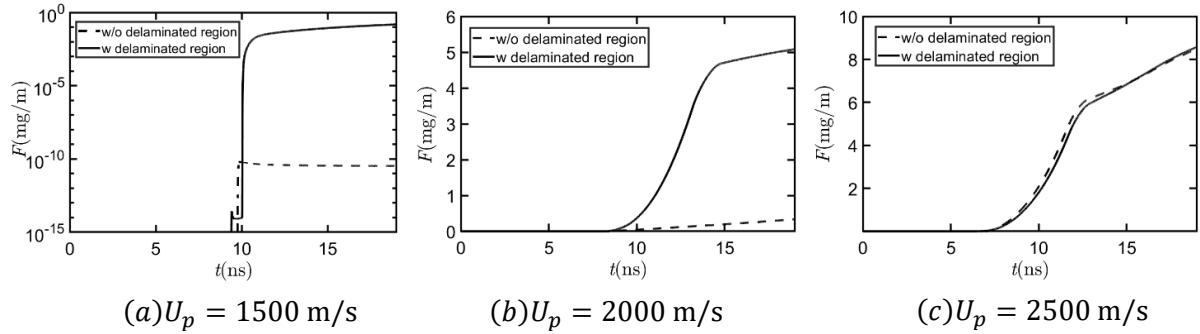

$(a) U_p = 1500$ m/s $\qquad$ $(b) U_p = 2000$ m/s $\qquad$ $(c) U_p = 2500$ m/s

Figure 11. Comparison of $F$ vs $t$ with and without a delaminated region (see Figures 3 (a) and (b) respectively). The response of the crystals with and without the delaminated region at the corner to $U_p = 1500$ m/s, 2000 m/s, and 2500 m/s shocks are compared in (a), (b), and (c) respectively.

accelerated into the void space, leading to collapse of the void. The contour plot of velocity magnitude in Figure 10 (f) shows that the delaminated binder interface reaches a velocity of nearly 4 km/s before impinging on the corner of the crystal. The resulting sudden post-impact increase in pressure and temperature is observed in Figure 10 (g) and (h), respectively. The augmented compression at the corner of the HMX crystal causes the temperature to reach ~4000 K, leading to ignition. Therefore, the presence of a delaminated region at the sharp corner significantly enhances the energy localization in HMX through shock focusing and facilitates the ignition of the crystal.

The effects of shock strength ($U_p$) on the ignition characteristics at a delamination region at the sharp corner is demonstrated in Figure 11 by plotting $F$ against $t$. Figure 11 (a) shows that for $U_p = 1500$ m/s the HMX crystal with delaminated region produces $F \sim 1$ mg/m after ignition. However, $F \sim O(10^{-10})$ in the calculation without the delaminated region is, i.e., the crystal without the delaminated regions at the corner does not ignite at $U_p = 1500$ m/s. Figure 11 (b) shows that both configurations, with and without the delaminated region, ignite when subjected to a $U_p = 2000$ m/s shock. However, the reaction rates of crystals with and without the delaminated region are quite different. $F$ of the crystal with the delaminated sharp corner reaches ~5 mg/m within 15 ns, compared to ~0.5 mg/m in the calculation without the delaminated sharp corner. At higher velocities, e.g. at $U_p = 2500$ m/s, the presence of the delaminated region is not found to significantly influence the burning rate of the HMX crystal with the sharp corner($\theta = 90^0$). Therefore, a delamination region at the sharp corner on the surface of a crystal can significantly enhance the sensitivity of a PBX sample by providing a potential site for initiation.

### 3.2.2 Semicircular delamination at the planar interface between crystal and binder

Delamination may also occur at the planar interfaces between the crystal and binder in PBXs. Therefore, the possibility of hotspot formation due to a delaminated region at a planar crystal-binder interface is investigated via the configuration shown in Figure 3(c). The semicircular void region has a diameter of 14 μm. The response of this configuration to a shock incident from the top boundary, corresponding to $U_p = 1500$ m/s, 2000 m/s, and 2500 m/s, is computed.

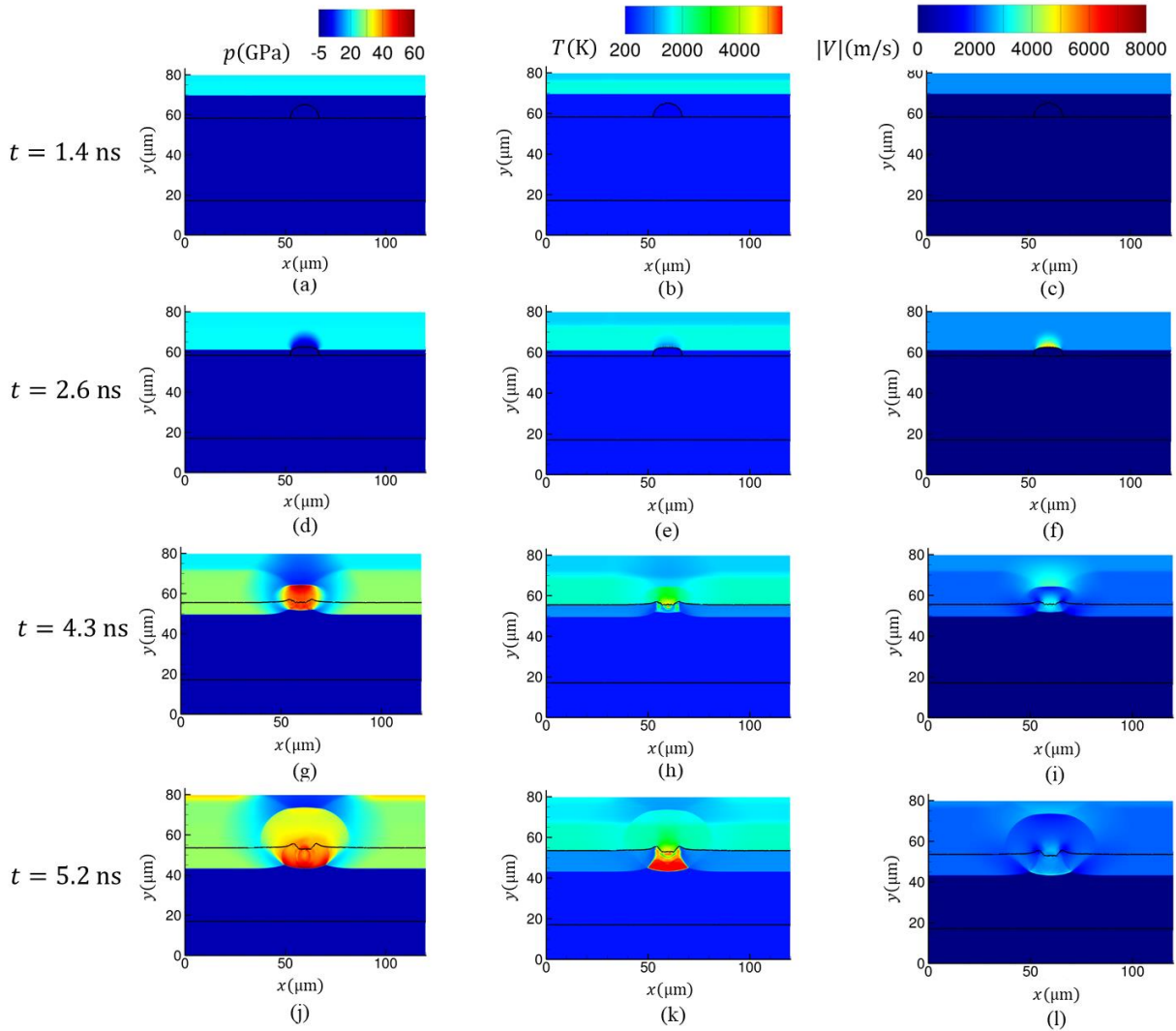

Figure 12. Sequence of pressure (left column), temperature (middle column), and velocity magnitude (right column) contours at four different instances during the $U_p = 2500$ m/s shock interaction with a semi-circular delamination at the crystal-binder interface.

The crystal in this configuration is found to ignite when subjected to a $U_\text{p} = 2500$ m/s shock. The mechanism of ignition due to shock induced collapse of the semicircular delamination region is shown through a sequence of contour plots of pressure $(p)$, temperature $(T)$, and velocity magnitude $(|V|)$ in Figure 12, plotted at four different time instances. Figure 12 (a), (b), and (c) show the contours of $p, T,$ and $|V|$ respectively, at t = 1.4 ns, as the incoming shock approaches the planar crystal-binder interface. The contours of $p, T,$ and $|V|$ shortly after arrival of the incoming shock at 2.6 ns are shown in Figure 12 (c), (d), and (e) respectively. Contours of $p$ and $T$, in Figure 12 (c) and (d) respectively, show that the incoming shock reflects as a rarefaction wave from the delaminated binder-void interface. Figure 12 (e) shows that the arrival of the incoming shock accelerates the binder-void interface to a velocity of ~5000 m/s. The upstream surface of the binder forms a jet that impinges on the HMX crystal at high speed as the delaminated region collapses. Localized high pressure (~ 50 GPa) and temperature (~ 4000 K) are observed at the surface of the HMX crystal upon closure of the delaminated region at 4.3 ns in Figure 12 (g) and (h), respectively. Figure 12 (h) shows that the temperature at the resulting hotspot is ~4000 K. Consequently, a self-sustained chemical reaction initiates in the crystal at the location of the initial delaminated crystal-binder interface. Figure 12 (k) shows that a sustained reaction front develops after the ignition of the HMX crystal. Therefore, shock interaction with the delaminated interface between the HMX crystal and binder at $U_p = 2500$ m/s leads to the formation of a critical hotspot.

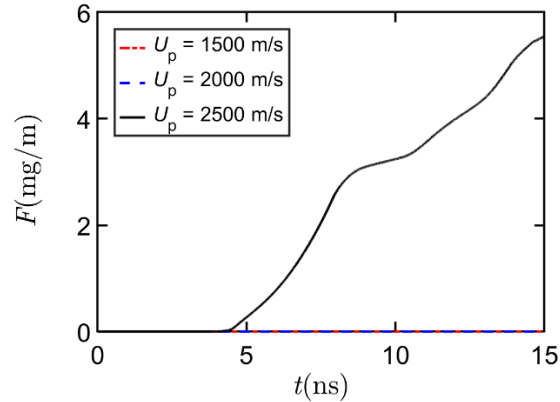

Figure 13. Comparison of F vs t during $U_p = 1500, 2000,$ and $2500$ m/s shock interaction with a semi-circular delamination region at the crystal-binder interface.

The total mass of combustion product formed in the cross-sectional area of the crystal ($F$) for shocks of $U_p = 1500$ m/s, 2000 m/s, and 2500 m/s are compared in Figure 13. Figure 13 shows that $F$ increases rapidly for the $U_p = 2500$ m/s shock. However, $F$ remains ~0 at $U_p = 1500$ m/s and 2000 m/s, i.e. initiation is not triggered at the two lower velocities.

### 3.3 Circular void in the binder near a planar surface of the crystal

Ignition of a HMX crystal due to shock-induced collapse of a circular void in the binder phase is studied in this section. A circular void, 14 μm in diameter, is located in the Estane binder, with its lower surface 1.4 μm away from a planar interface of an HMX crystal. The initial computational setup for this calculation is shown in Figure 3 (d). A supported shock is applied at the north boundary of the computational domain, at the three different shock strengths $(U_p = 1500$ m/s, 2000 m/s, and 2500 m/s$)$ to study if the energy localized near the crystal-binder interface is sufficient to ignite the crystal.

The response of the configuration shown in Figure 3 (d) to a $U_p = 2500$ m/s shock is shown through sequences of pressure$(p)$, temperature$(T)$, and velocity magnitude$(|V|)$ contours at four different time instances in Figure 14. The left, middle and right columns in Figure 14 show $p$, $T$, and $|V|$ contours respectively. The four rows in Figure 14 from top to bottom show four different time instances (in the order: t = 2.7, 4.3, 5.7 and 7.5 ns). Figure 14 (a) shows that the void in the binder begins to collapse as the incoming shock arrives at the north end of the binder-void interface. Contours of $p$ and $T$ in Figure 14 (a) and (b) respectively show that the incident shock reflects as a rarefaction wave from the binder-void interface. Figure 14 (d), (e), and (f) respectively show that a high-speed re-entrant jet forms as the void in the binder collapses. The velocity magnitude contour in Figure 14 (f) shows that the binder-void interface at the top accelerates to ~6000 m/s under the action of the incoming shock. Figure 14 (g) and (h) show the intense pressure and temperature localization at the location of void collapse. The energy localization in the binder after the shock-induced collapse of the void sets off a strong blast wave. Figure 14 (g), (h), and (i) show that the blast wave travels across the crystal-binder interface. The sudden compression of HMX near the interface due to the interaction with the blast wave increases the local temperature and sets off self-sustained chemical reactions. The pressure and temperature contours in Figure

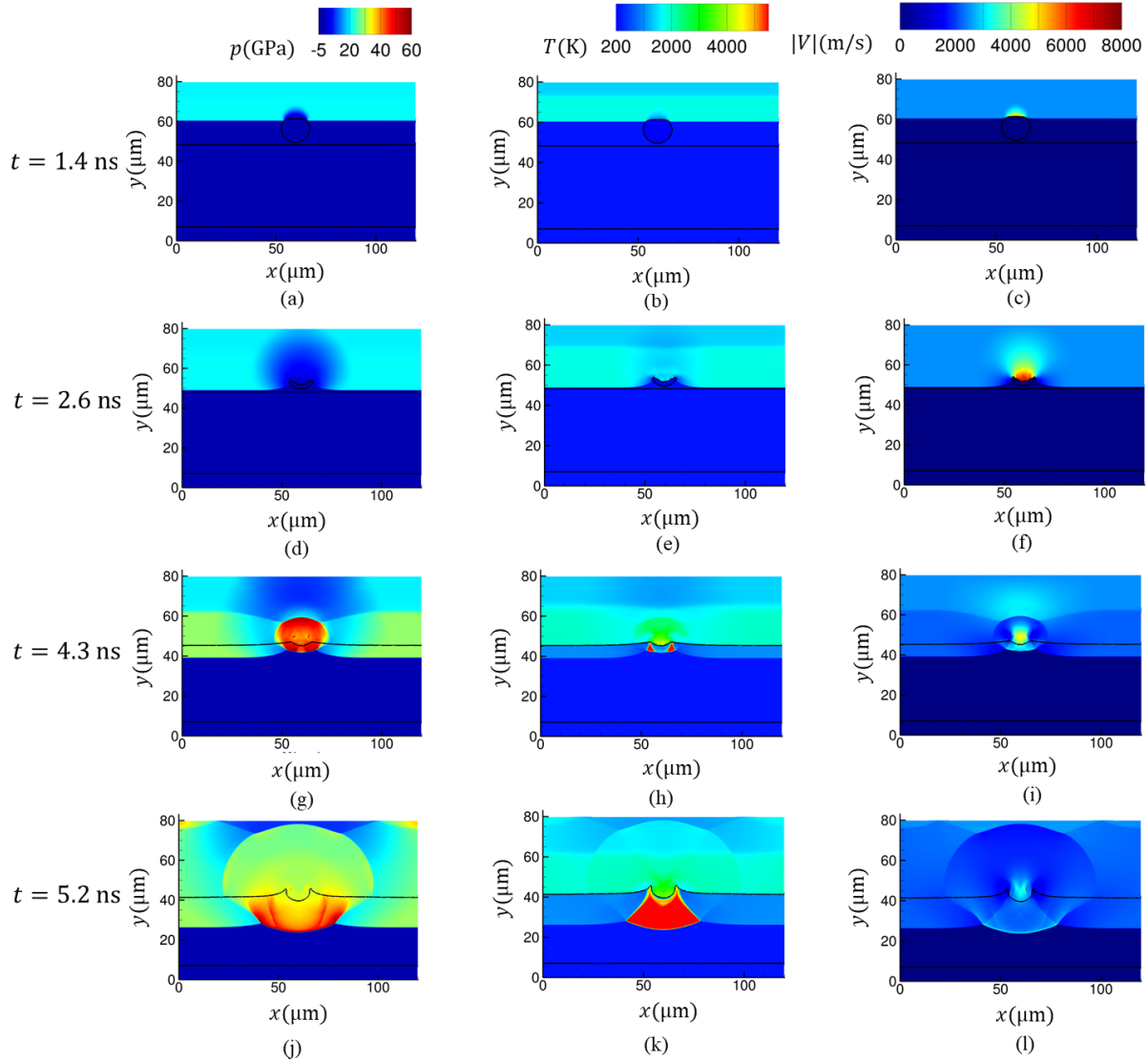

Figure 14. Sequence of contours of pressure (left column), temperature (middle column), and velocity magnitude (right column) at four different instances during the $U_p = 2500$ m/s shock interaction with a void (14 μm in diameter) in the binder located near the crystal-binder interface.

14 (j) and (k) show that the reaction front in HMX remains coupled with the blast wave transmitted into the HMX and travels as a strong reaction wave. The time evolution of the total mass of the reaction product plotted in Figure 15 supports the observation of sustained reaction in HMX crystal in Figure 14 (d). Figure 15 shows that the HMX crystal reacts at a high rate under the action of the blast wave that originates from the void-collapse site in the binder. Therefore, at $U_p = 2500$ m/s, the shock-induced collapse of the void in binder ignites the adjacent HMX crystal. Figure 15 also shows that in this case the HMX crystal ignites at $U_p = 2000$ m/s. However, the $U_p = 1500$ m/s shock fails to initiate self-sustained reaction in the HMX crystal in this case as well. Thus, in comparison to shock focusing at the sharp corners with delaminated crystal-binder interface (section 3.2.1), the extra-crystal void collapse (i.e. for a void partially or wholly in the binder) near a planar crystal-binder interface appears to be a weaker mechanism for initiation, at least for the particular range of conditions investigated here.

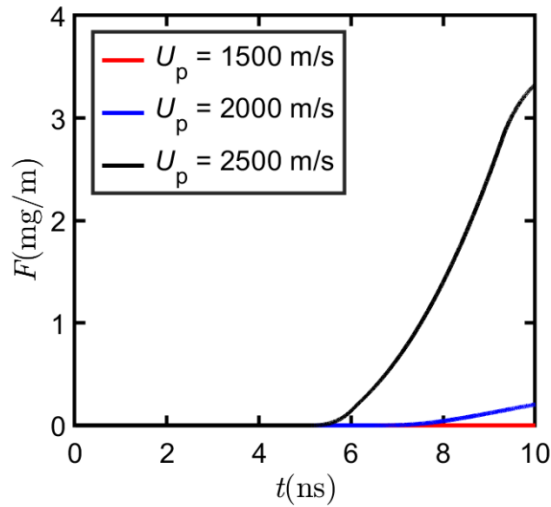

Figure 15. Comparison of F vs t for $U_p$ = 1500, 2000, and 2500 m/s, shock interaction with a void (14 µm in diameter, centered at 15.4 µm away from the interface) in the binder located near the crystal-binder interface.

### 3.4 The effect of a void in the HMX crystal on its ignition characteristics under shock loading

The grains of crystalline energetic materials used in PBXs often have internal defects such as cracks and voids. The contribution to localization and relative importance (in comparison to the mechanisms of shock focusing at sharp corners and extra-crystalline voids examined above) of intra-crystalline voids is examined in this section. We study the response of an HMX crystal with internal defects to incoming shocks of $U_p$ = 1500 m/s and 2000 m/s. The 2D section from a CT scan of the HMX crystal shown in Figure 1 is used in this calculation. The crystal geometry is imported from the image into the computational domain using the algorithm described in section 2.5. Figure 3 (e) shows the initial computational setup for the calculation of shock interaction with the crystal. The crystal-binder configuration shown in Figure 3 (e) is subjected to shocks travelling downward from the north boundary of the computation domain. The results obtained from these calculations are compared with corresponding calculations of the crystal without the internal void. These comparisons show the effect of the void within the crystal on its sensitivity under shock loading.

The response of the crystal without and with the internal void under the action of the $U_p$ = 1500 m/s shock is shown through a sequence of temperature contours in the left and right column of Figure 16, respectively. Figure 16 (a) and (b) show the temperature contours at an instant ($t$ = 5 ns) when the shock has already been transmitted into the crystal from the binder. The bulk temperature in the crystal increases behind the transmitted shock. Note that the bulk heating of the HMX crystal behind the transmitted shock by itself is not sufficient for ignition. In this geometric configuration, the only relevant mechanisms for energy localization within the HMX crystal are through shock focusing at the sharp corners of the crystal (section 3.1) or through collapse of the void within the crystal [34]. Figure 16 (c) and (d) show that in the current case the HMX crystal does not ignite at the sharp concave corners; as seen in section 3.1, at $U_p$ = 1500 m/s the incident shock is not strong enough to localize energy to trigger initiation at the corners. However, the temperature contours in Figure 16 (d) show that the collapse of the void within the crystal initiates a local hotspot. Figure 16 (f) shows that the hotspot grows into a self-sustained reaction zone. Therefore, even at the lowest shock strength, i.e., for $U_p$ = 1500 m/s, shock-induced collapse of the intra-crystal void localizes sufficient energy for ignition. On the other hand, the reaction product mass produced in the HMX crystal without the internal void remains close to zero over the period of the calculation as shown in Figure 17 (a).

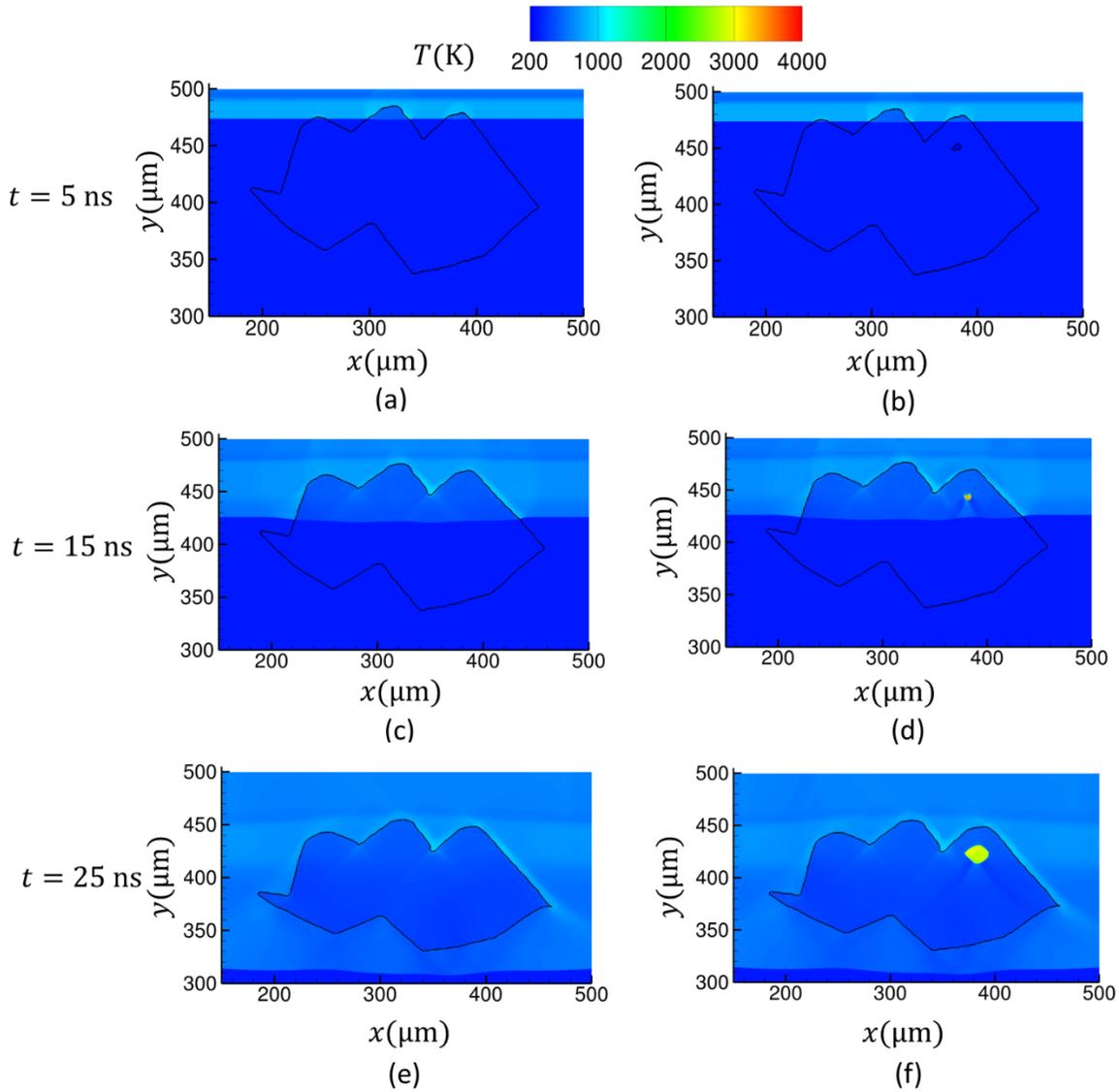

Figure 16. Sequence of temperature contours during a $U_p = 1500$ m/s shock interaction with an imaged geometry of HMX crystal embedded in Estane binder.

The thermal responses of the HMX crystals without and with an internal void under the action of $U_p = 2300$ m/s shock are compared through a sequence of temperature contours in Figure 18. Figure 18 (a) and (b) show that the HMX crystals do not ignite due to the bulk heating of the crystals under the action of the transmitted shock. However, Figure 18 (c) and (d) show that both HMX crystals ignite at the sharp concave corners under the action of $U_p = 2300$ m/s shock. Furthermore, the temperature contour in Figure 18 (d) shows that the HMX crystal with the defect ignites at the location of the internal void after the void collapses. Figure 18 (e) and (f) show that both HMX crystals are almost fully consumed by the reaction at 25 ns. However, the comparison of the mass of reaction products in Figure 17 (b) shows that the crystal with the internal void burns at a higher rate than the crystal without the internal void. Therefore, the presence of the internal void in the HMX crystal increases the amount of energy localized in the crystal under shock loading and consequently enhances its reactivity and sensitivity to incoming shocks.

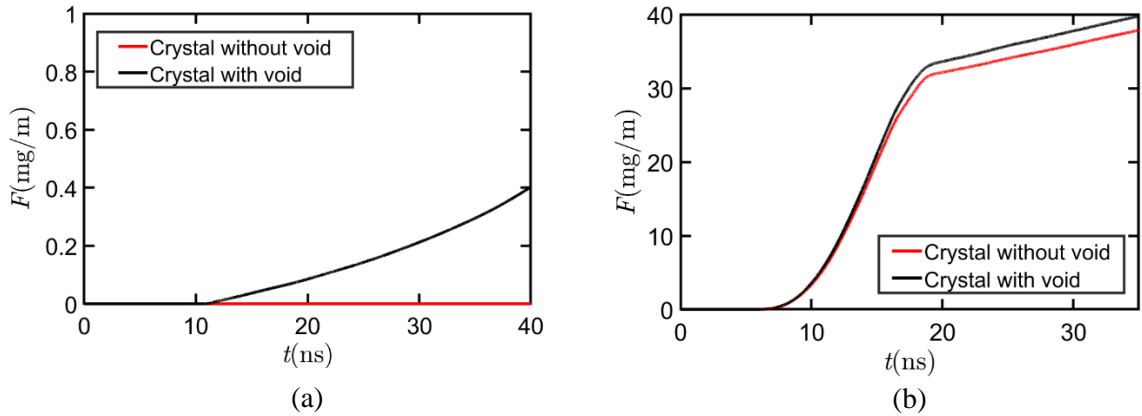

Figure 17. The comparison of *F* vs *t* during interaction of the shock with HMX crystal, for a crystal with and without a void at (a) $U_p = 1500$ m/s and (b) $U_p = 2300$ m/s.

## 4. Conclusions

Energetic crystals of size ~10-100 μm embedded in a polymeric binder are the building blocks of PBXs. Here, the grain-scale thermo-mechanics of shock initiation is studied through high-resolution reactive simulations. It is shown that the microstructural features of these energetic crystals and the nature of their interfaces with the binder contribute to energy localization and the formation of intense hotspots under shock loading. Specifically, the present simulations elucidate the mechanistic details of energy localization and the possibility of ignition due to three types of microstructural features: 1) sharp crystal corners, 2) delaminated regions at the crystal-binder interfaces, and 3) intra-crystal voids.

The sharp concave corners of an irregular shaped HMX crystal are found to be potential locations of energy localization under shock loading. Compression due to shock-focusing increases the temperature at these sharp concave corners crystals and leads to the development of local hotspots. This may explain the mechanism for hotspot formation at the corners and edges of HMX crystals observed by Johnson et al. [10] in their experimental study. However, the ignition of HMX crystals due to shock-focusing at sharp corners is found to be strongly dependent on the strength of the incoming shock and the geometry of the sharp corner. Another potential mechanism for energy localization and ignition of HMX crystals at the crystal-binder interface is through collapse of extra-crystal voids due to delaminated regions. The collapse of the delaminated region at the sharp corners further enhances the localization of thermal energy due to shock focusing. Furthermore, the results show that collapse of delaminated regions at the planar crystal-binder interface under the action of strong shocks can also lead to ignition of the crystal. Collapse of voids in the binder is found to ignite nearby HMX crystals, due to strong blast waves which travel from the binder into the crystal. The local heating of the HMX crystals due to the adiabatic compression under the action of these blast waves is sufficient to initiate self-sustained reaction in the crystals. Defects within HMX crystals in PBXs are also found to contribute to critical hotspot formation under shock loading.

The current results show that several microstructural features of PBXs, such as sharp crystal corners, delamination at the crystal-binder interface, voids in the binder, and internal defects within the crystals all contribute to energy localization under the action of mechanical insults such as shocks. However, the extent and intensity of energy localization can depend on the topological/morphological parameters defining these microstructural features (such as the angle of the sharp corners or the shape and size of the delamination regions and voids), strength of the shock, and thermo-mechanical properties of the crystal and binder. For the range of parameters and configurations studied here, intra-crystalline voids, sharp corners and extra-crystalline voids, ranked in the decreasing order of sensitivity, serve as sites for initiation. Highly intense hotspots arise due to void collapse, while less intense, but critical hotpots can be produced at surface features (asperities, debonded zones) as well. These results partially explain the observed hotspot locations and differences in intensities reported in the experimental studies of Johnson *et al*. [10].

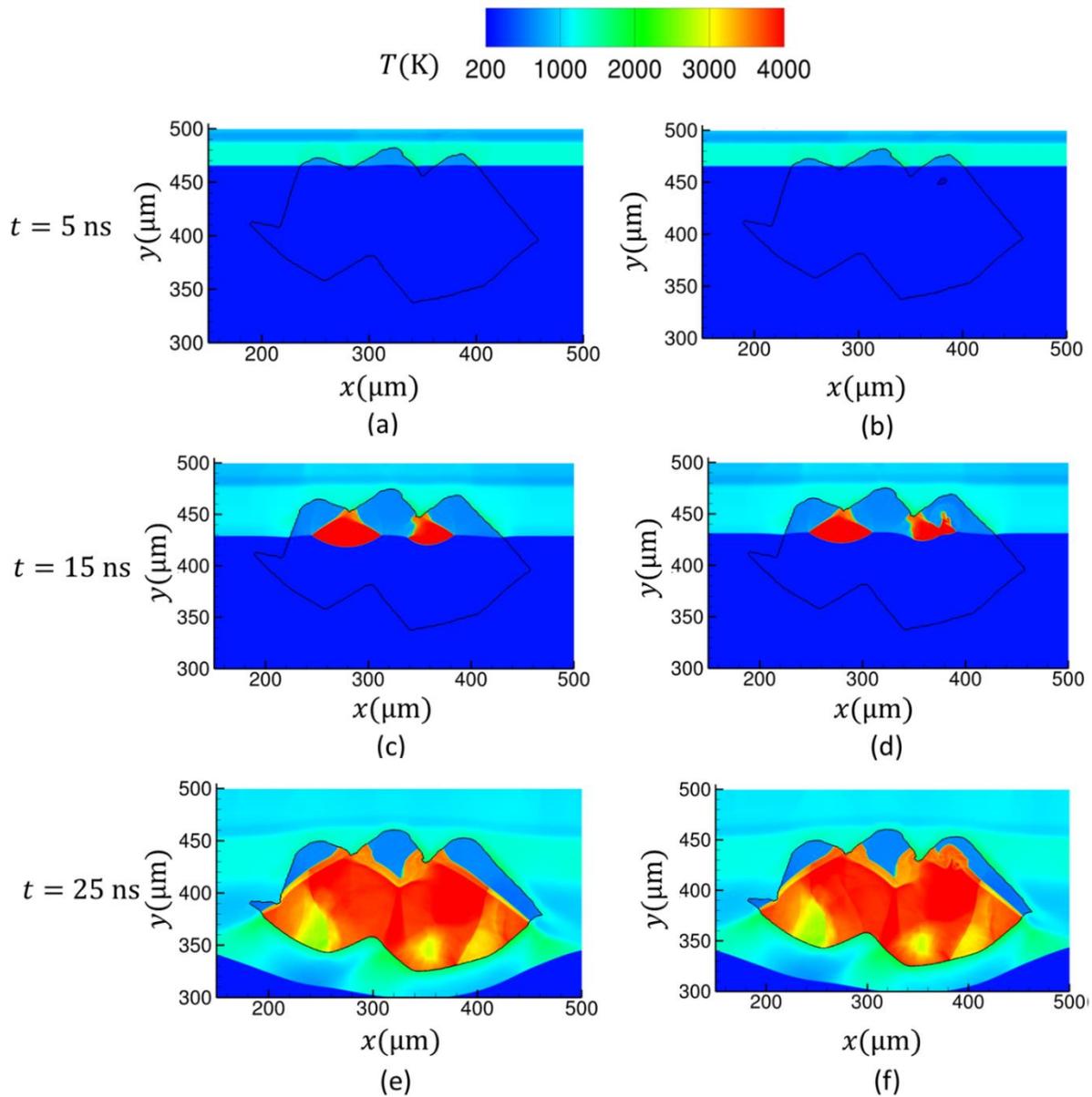

Figure 28. Sequence of temperature contours during a $U_p = 2300$ m/s shock interaction with an imaged geometry of a HMX crystal embedded in Estane binder.

Further studies to characterize the relationship between the morphological parameters and the ignition characteristics of PBXs are underway. The simplified 2D configurations used in the current work provide computationally tractable surrogates for studying the energy localization mechanisms due to the microstructural features seen in PBXs. However, 3D calculation will be necessary for predictive modeling of shock-induced initiation of PBXs. Such 3D calculations of shock-induced energy localization in realistic HMX crystals in binders are begin pursued in ongoing work.

**Acknowledgments**

This work was supported by an AFOSR-MURI grant (Grant number: FA9550-19-1-0318; program manager: Martin Schmidt). The authors acknowledge illuminating discussions with Dana D. Dlott (Department of Chemistry, University of Illinois Urbana-Champaign), Xuan Zhou (Department of Chemistry, University of Illinois Urbana-Champaign), and Belinda P. Johnson (Department of Chemistry, University of Illinois Urbana-Champaign). The authors would like to thank Xuan Zhou (Department of Chemistry, University of Illinois Urbana-Champaign) for providing the high-resolution CT image of the HMX crystal in binder used for numerical simulations in this work.

**Appendix**

**A1. The three-step reaction model for HMX**

The thermal decomposition of HMX is modeled using a three step chemical kinetic model involving four groups of reaction products[6]. The three steps are given as,

$$\text{HMX } (C_4H_8N_8O_8) \rightarrow \text{fragments } (CH_2NNO_2) \tag{A1}$$

$$\text{fragments } (CH_2NNO_2) \rightarrow \text{intermediate gases } (CH_2O, N_2O, HCN, HNO_2) \tag{A2}$$

$$2 \times \text{intermediate gases } (CH_2O, N_2O, HCN, HNO_2) \rightarrow \text{final gases } (N_2, H_2O, CO_2, CO) \tag{A3}$$

The rate equations for all the species are given as,

$$\dot{Y}_1 = -Y_1 Z_1 e^{-\frac{E_1}{RT}} \tag{A4}$$

$$\dot{Y}_2 = Y_1 Z_1 e^{-\frac{E_1}{RT}} - Y_2 Z_2 e^{-\frac{E_2}{RT}} \tag{A5}$$

$$\dot{Y}_3 = Y_2 Z_2 e^{-\frac{E_2}{RT}} - Y_3^2 Z_3 e^{-\frac{E_3}{RT}} \tag{A6}$$

$$\dot{Y}_4 = Y_3^2 Z_3 e^{-\frac{E_3}{RT}} \tag{A7}$$

where, $Y_i$ is the mass fraction of the $i^{\text{th}}$ species. $Z_j$ and $E_j$ are the frequency factor and activation energy for the $j^{\text{th}}$ reaction. $R$ is the universal gas constant and $T$ is the temperature. The values for each of these constants are given in Table A1 [6].

The total heat release rate $(\dot{Q}_R)$ from all the reactions (Eq. (7)) and is given as,

$$\dot{Q}_R = \sum_{j=1}^{3} Q_j \dot{Y}_j \tag{A8}$$

where, $j = 1 - 3$ is the reaction number (Eq. (A1) – (A3)), $Q_j$ is the energy release from each of the reactions. The value for $Q_j$ is tabulated in Table A1.

The values of $k$ and $c_P$ for the reaction mixture(Eq. (7)) are obtained by weighted mass fraction average of the specific heat and thermal conductivity for the four species,

$$c_P = \sum_{i=1}^{4} c_P^i Y_i \tag{A9}$$

$$k = \sum_{i=1}^{4} k_i Y_i \tag{A10}$$

where, $c_P^i$ and $k_i$ are the specific heat capacity and thermal conductivity for the four species obtained from the work of Tarver et al. [6]. The values for the $c_P^i$ and $k_i$ for each of the species for various temperatures are listed in Table A2 [6].

| Property | Value |
| --- | --- |
| $\ln Z_1$ (s$^{-1}$) | 48.7 |
| $\ln Z_2$ (s$^{-1}$) | 37.3 |
| $\ln Z_3$ (s$^{-1}$) | 28.1 |
| $E_1$(kcal/m) | 52.7 |
| $E_2$(kcal/m) | 44.1 |
| $E_3$(kcal/m) | 34.1 |
| $Q_1$(cal/g) at 298 K | +100 |
| $Q_2$(cal/g) at 298 K | -300 |
| $Q_3$(cal/g) at 298 K | -1200 |

Table A1. HMX chemical reaction parameters for Tarver 3-equation model[6].

| Property | Temperature | HMX | Fragments | Intermediate Gases | Final Gases |
|---|---|---|---|---|---|
| | 293 K | 0.24 | 0.22 | 0.24 | 0.27 |
| | 433 K | 0.34 | 0.31 | 0.27 | 0.28 |
| | 533 K | 0.40 | 0.36 | 0.29 | 0.29 |
| Specific Heat, $c_P$ (cal/(g.K)) | 623 K | 0.46 | 0.42 | 0.31 | 0.30 |
| | 773 K | 0.55 | 0.50 | 0.35 | 0.31 |
| | > 1273 K | 0.55 | 0.50 | 0.42 | 0.35 |
| | 293 K | $1.23 \times 10^{-3}$ | $6.5 \times 10^{-4}$ | $1 \times 10^{-4}$ | $1 \times 10^{-4}$ |
| Thermal Conductivity, $k$ (cal/(cm.s.K)) | 433 K | $9.7 \times 10^{-4}$ | $5.0 \times 10^{-4}$ | $1 \times 10^{-4}$ | $1 \times 10^{-4}$ |
| | 533 K | $8.1 \times 10^{-4}$ | $4.0 \times 10^{-4}$ | $1 \times 10^{-4}$ | $1 \times 10^{-4}$ |
| | > 623 K | $7.0 \times 10^{-4}$ | $3.0 \times 10^{-4}$ | $1 \times 10^{-4}$ | $1 \times 10^{-4}$ |

Table A2. Specific heat and thermal conductivity for all the four species from the Tarver 3-equation [6] model for various temperatures.